\documentclass[journal]{IEEEtran}
\usepackage{amsmath}
\usepackage[pdftex]{graphicx}
\usepackage{array}
\usepackage[printonlyused]{acronym}
\usepackage{cite}
\usepackage{multirow}

\title{Knowledge-Defined Networking}

\author{\IEEEauthorblockN{Albert Mestres\IEEEauthorrefmark{1},
Alberto Rodriguez-Natal\IEEEauthorrefmark{1}, 
Josep Carner\IEEEauthorrefmark{1}, 
Pere Barlet-Ros\IEEEauthorrefmark{1}, 
Eduard Alarc\'on\IEEEauthorrefmark{1}, \\
Marc Sol\'e\IEEEauthorrefmark{2},
Victor Munt\'es-Mulero\IEEEauthorrefmark{2}, 
David Meyer\IEEEauthorrefmark{3},
Sharon Barkai\IEEEauthorrefmark{4}, 
Mike J Hibbett\IEEEauthorrefmark{5}, 
Giovani Estrada\IEEEauthorrefmark{5}, \\ 
Khaldun Ma`ruf\IEEEauthorrefmark{6}, 
Florin Coras\IEEEauthorrefmark{7},
Vina Ermagan\IEEEauthorrefmark{7},
Hugo Latapie\IEEEauthorrefmark{7},
Chris Cassar\IEEEauthorrefmark{7},
John Evans\IEEEauthorrefmark{7},
Fabio Maino\IEEEauthorrefmark{7}, \\
Jean Walrand\IEEEauthorrefmark{8} and
Albert Cabellos\IEEEauthorrefmark{1}}

\IEEEauthorblockA{ \IEEEauthorrefmark{1} Universitat Polit\`ecnica de Catalunya}
\IEEEauthorblockA{\IEEEauthorrefmark{2} CA Technologies  }
\IEEEauthorblockA{\IEEEauthorrefmark{3} Brocade Communication  }
\IEEEauthorblockA{\IEEEauthorrefmark{4} Hewlett Packard Enterprise \\ }
\IEEEauthorblockA{\IEEEauthorrefmark{5} Intel R\&D   }
\IEEEauthorblockA{\IEEEauthorrefmark{6} NTT Communications }
\IEEEauthorblockA{\IEEEauthorrefmark{7} Cisco Systems  }
\IEEEauthorblockA{\IEEEauthorrefmark{8} University of California, Berkeley }

}

\begin{document}

\maketitle

\begin{abstract}

The research community has considered in the past the application of Artificial Intelligence (AI) techniques to control and operate networks. A notable example is the Knowledge Plane proposed by D.Clark et al. However, such techniques have not been extensively prototyped or deployed in the field yet. In this paper, we explore the reasons for the lack of adoption and posit that the rise of two recent paradigms: Software-Defined Networking (SDN) and Network Analytics (NA), will facilitate the adoption of AI techniques in the context of network operation and control. We describe a new paradigm that accommodates and exploits SDN, NA and AI, and provide use cases that illustrate its applicability and benefits. We also present simple experimental results that support its feasibility. We refer to this new paradigm as Knowledge-Defined Networking (KDN).

%
%
\end{abstract}


\acrodef{SDN}{Software-Defined Networking}
\acrodef{ANN}{Artificial Neural Network}
\acrodef{DC}{Data Center}
\acrodef{VM}{Virtual Machine}
\acrodef{NFV}{Network Function Virtualization}
\acrodef{VNF}{Virtual Network Function}
\acrodef{ML}{Machine Learning}
\acrodef{KP}{Knowledge Plane}
\acrodef{KDN}{Knowledge-Defined Networking}
\acrodef{NA}{Network Analytics}
\acrodef{}{}
\acrodef{}{}

\section{Introduction}
\label{sec:intro}

D. Clark et al. proposed ``A Knowledge Plane for the Internet''~\cite{clark}, a new construct that relies on \ac{ML} and cognitive techniques to operate the network. A \ac{KP} would bring many advantages to networking, such as automation (recognize-act) and recommendation (recognize-explain-suggest), and it has the potential to represent a paradigm shift on the way we operate, optimize and troubleshoot data networks. However, at the time of this writing, we are yet to see the \ac{KP} prototyped or deployed. Why?

One of the biggest challenges when applying \ac{ML} for network operation and control is that networks are inherently distributed systems, where each node (i.e., switch, router) has only a partial view and control over the complete system.
Learning from nodes that can only view and act over a small portion of the system is very complex, particularly if the end goal is to exercise control beyond the local domain. The emerging trend towards centralization of control will ease this flavor of complexity. In particular, the \ac{SDN} paradigm~\cite{sdn} decouples control from the data plane and provides a logically centralized control plane, i.e. a logical single point in the network with knowledge of the whole.


In addition to the ''softwarization'' of the network, current network data plane elements, such as routers and switches, are equipped with improved computing and storage capabilities. This has enabled a new breed of network monitoring techniques, commonly referred to as network telemetry~\cite{telemetry}. Such techniques provide real-time packet and flow-granularity information, as well as configuration and network state monitoring data, to a centralized \ac{NA} platform~\cite{analytics}. In this context, telemetry and analytics technologies provide a richer view of the network compared to what was possible with conventional network management approaches.

In this paper, we advocate that the centralized control offered by \ac{SDN}, combined with a rich centralized view of the network provided by network analytics, enable the deployment of the \ac{KP} concept proposed in~\cite{clark}. In this context, the \ac{KP} can use \ac{ML} and Deep Learning (DL) techniques to gather knowledge about the network, and exploit that knowledge to control the network using logically centralized control capabilities provided by \ac{SDN}. We refer to the paradigm resulting from combining \ac{SDN}, telemetry, Network Analytics, and the Knowledge Plane as Knowledge-Defined Networking.

This paper first describes the \ac{KDN} paradigm and how it operates. Then, it describes a set of relevant use-cases that show the applicability of such paradigm to networking and the benefits associated with using \ac{ML}. In addition, for some use-cases, we also provide early experimental results that show their feasibility. We conclude the paper by analyzing the open research challenges associated with the \ac{KDN} paradigm.

\section{A Knowledge Plane for SDN Architectures}
\label{sec:kp}

\begin{figure}
  \centering
  \includegraphics[width=0.94\columnwidth]{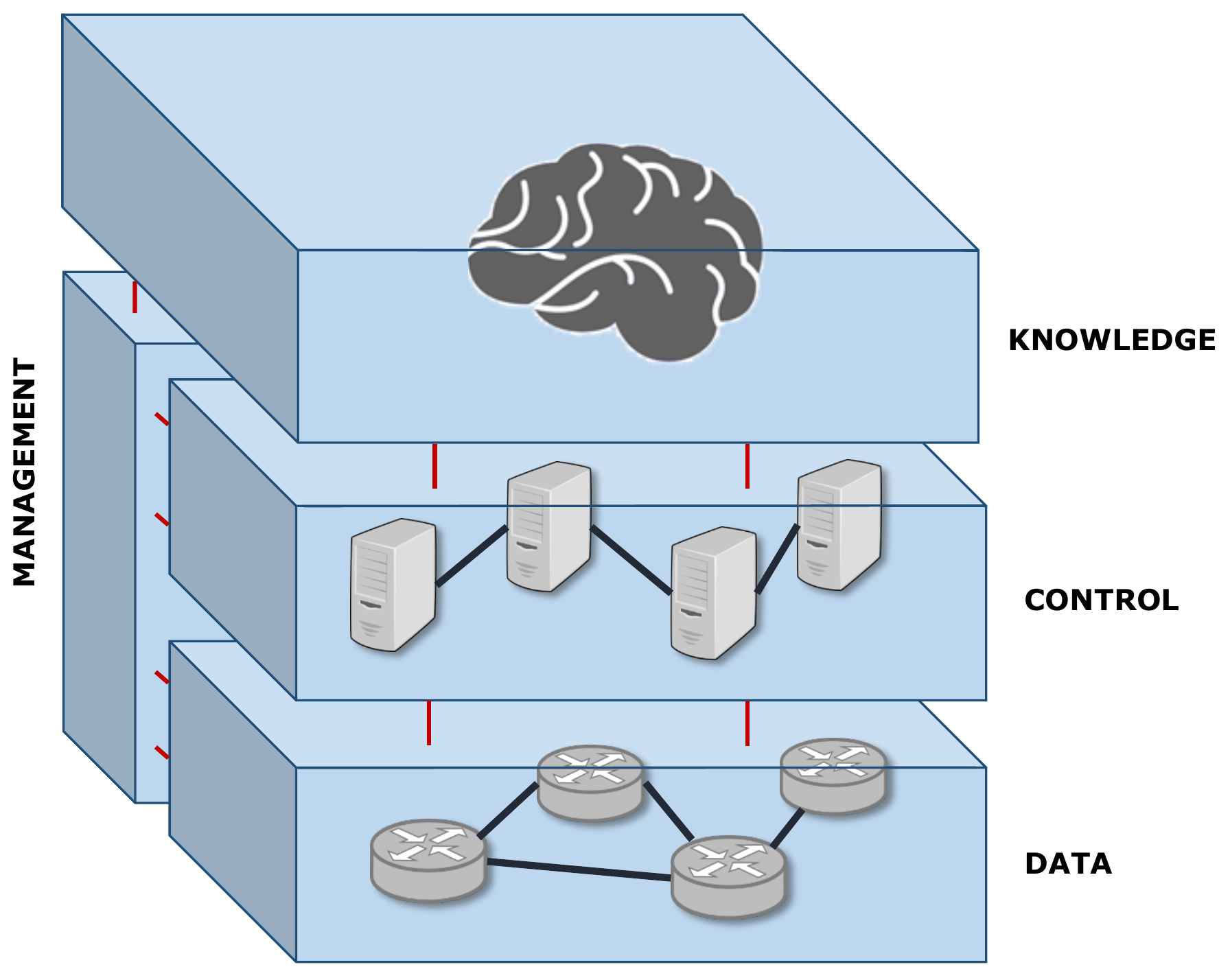}
  \caption{KDN planes}
  \label{fig:planes}
\end{figure}

This paper restates the concept of \acf{KP} as defined by D. Clark et al.~\cite{clark} in the context of \ac{SDN} architectures. The addition of a \ac{KP} to the traditional three planes of the \ac{SDN} paradigm results in what we call Knowledge-Defined Networking. Fig.~\ref{fig:planes} shows an overview of the \ac{KDN} paradigm and its functional planes.

The \textit{Data Plane} is responsible for storing, forwarding and processing data packets. In \ac{SDN} networks, data plane elements are typically network devices composed of line-rate programmable forwarding hardware. They operate unaware of the rest of the network and rely on the other planes to populate their forwarding tables and update their configuration. 

The \textit{Control Plane} exchanges operational state in order to update the data plane matching and processing rules. In an \ac{SDN} network, this role is assigned to the --logically centralized-- \ac{SDN} controller that programs \ac{SDN} data plane forwarding elements via a southbound interface, typically using an imperative language. While the data plane operates at packet time scales, the control plane is slower and typically operates at flow time scales.

The \textit{Management Plane} ensures the correct operation and performance of the network in the long term. It defines the network topology and handles the provision and configuration of network devices. In \ac{SDN} this is usually handled by the \ac{SDN} controller as well. The management plane is also responsible for monitoring the network to provide critical network analytics. To this end, it collects telemetry information from the data plane while keeping a historical record of the network state and events. The management plane is orthogonal to the control and data planes, and typically operates at larger time-scales. 

The \textit{Knowledge Plane}, as originally proposed by Clark, is redefined in this paper under the terms of \ac{SDN} as follows: \emph{the heart of the knowledge plane is its ability to integrate behavioral models and reasoning processes oriented to decision making into an \ac{SDN} network}. In the \ac{KDN} paradigm, the \ac{KP} takes advantage of the control and management planes to obtain a rich view and control over the network. It is responsible for learning the behavior of the network and, in some cases, automatically operate the network accordingly. Fundamentally, the \ac{KP} processes the network analytics collected by the management plane, transforms them into knowledge via \ac{ML}, and uses that knowledge to make decisions (either automatically or through human intervention). While parsing the information and learning from it is typically a slow process, using such knowledge automatically can be done at a time-scales close to those of the control and management planes.

\section{Knowledge-Defined Networking}
\label{sec:cognitive}

The Knowledge-Defined Networking paradigm operates by means of a control loop to provide automation, recommendation, optimization, validation and estimation. Conceptually, the KDN paradigm borrows many ideas from other areas, notably from black-box optimization~\cite{blackBoxOpt}, neural-networks in feedback control systems~\cite{NNControl1} and autonomic self-* architectures~\cite{selfArch}. In addition, recent proposals share the same vision stated in this paper~\cite{cobanets}. Fig. 2 shows the basic steps of the main KDN control. In what follows we describe these steps in detail.

\begin{figure}
  \centering
  \includegraphics[width=\columnwidth]{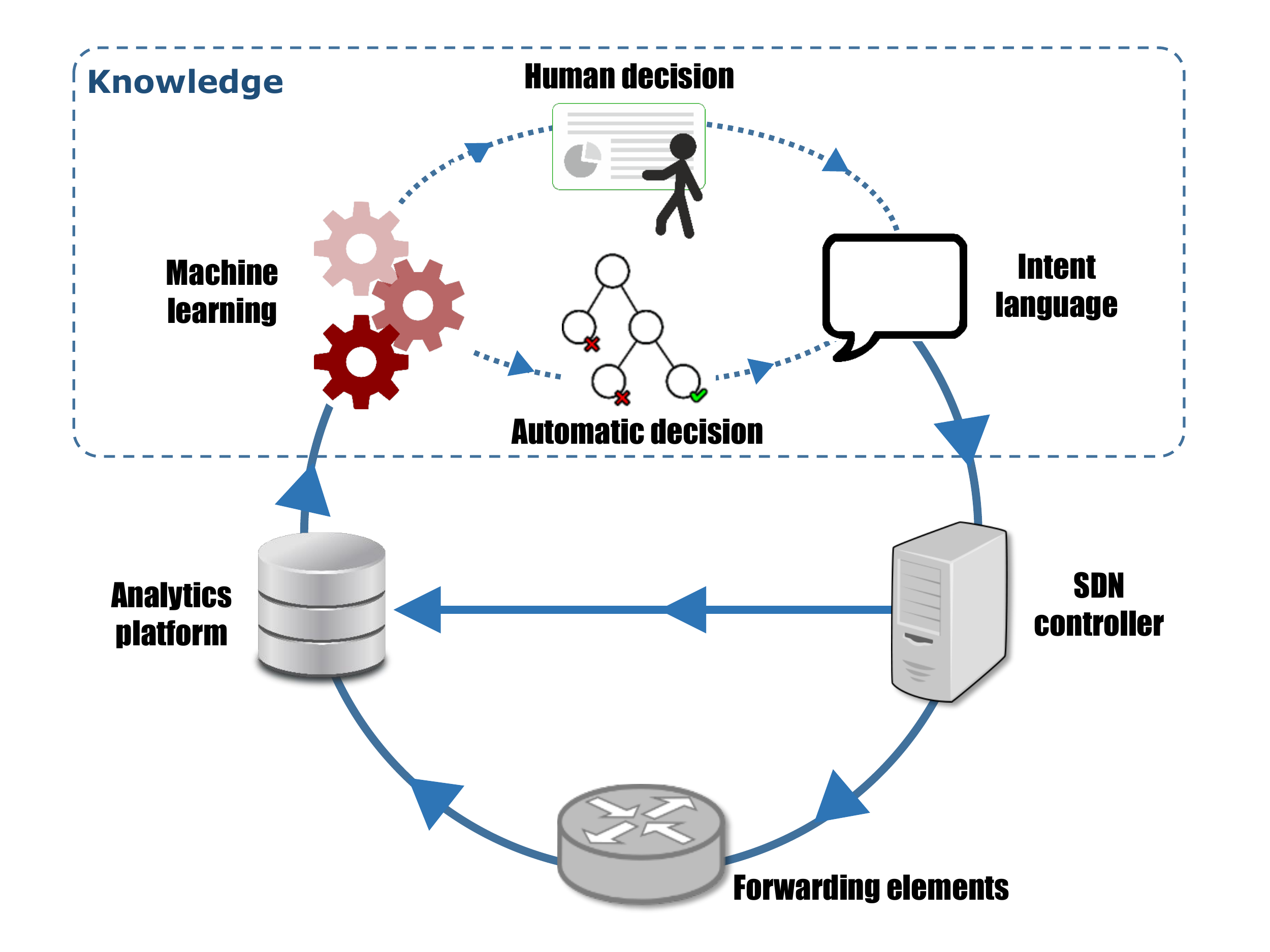}
  \caption{KDN operational loop}
  \label{fig:loop}
\end{figure}

\paragraph{Forwarding Elements \& SDN Controller $\rightarrow$ Analytics Platform}

The \textit{Analytics Platform} aims to gather enough information to offer a complete view of the network. To that end, it monitors the data plane elements in real time while they forward packets in order to access fine-grained traffic information. In addition, it queries the \ac{SDN} controller to obtain control and management state. The analytics platform relies on protocols, such as NETCONF (RFC 6241), NetFlow (RFC 3954) and IPFIX (RFC 7011), to obtain configuration information, operational state and traffic data from the network. The most relevant data collected by the analytics platform is summarized below. 

\begin{itemize}
\item Packet-level and flow-level data: This includes DPI information, flow granularity data and relevant traffic features.
\item Network state: This includes the physical, topological and logical configuration of the network.
\item Control \& management state: This includes all the information included both in the SDN controller and management infrastructure, including policy, virtual topologies, application-related information, etc. 
\item Service-level telemetry: In some scenarios the analytics platform will also monitor service-level information (e.g., load of the services, QoE, etc). This is relevant to learn the behavior of the application or service,  and its relation with the network performance, load and configuration.

\item External information: In some scenarios it may be useful to also have access to external information, such as social networks (e.g., amount of people attending a sports event), weather forecasts, etc. that may have a strong impact on the network.
\end{itemize}

In order to effectively learn the network behavior, besides having a rich view of the network, it is critical to observe as many different situations as possible. As we discuss in Section \ref{sec:challenges}, this includes different loads, configurations and services. To that end, the analytics platform keeps a historical record of the collected data.

\paragraph{Analytics Platform $\rightarrow$ Machine Learning}

The heart of the \ac{KP} are the \ac{ML} and Deep Learning algorithms, able to learn from the network behavior. The current and historical data provided by the analytics platform are used to feed learning algorithms that learn from the network and generate knowledge (e.g., a model of the network). We consider three approaches:

\begin{itemize}
\item \textit{Supervised learning:} The \ac{KP} learns a model that describes the behavior of the network, i.e. a function that relates relevant network variables to the operation of the network. The obtained model reflects correlations on the network behavior that are of interest to the network operator (e.g., the performance of the network as a function of the traffic load and network configuration). Supervised learning requires labeled training data and some feature engineering (e.g., feature construction and feature selection) to decide the relevant features prior to using them in \ac{ML} algorithms.
\item \textit{Unsupervised learning:} It is a data-driven knowledge discovery approach that can automatically infer a function that describes the structure of the analyzed data. Unsupervised learning does not require previously labeled samples and can highlight correlations in the data that the network operator may be unaware of. As an example, the \ac{KP} may be able to discover how the local weather affects the link's utilization. 
\item \textit{Reinforcement learning (RL):} In this approach a software agent aims to discover which actions lead to an optimal configuration. Formally in RL, the environment is typically modeled as a stochastic finite state machine where the agent sends inputs (actions) and receives outputs (observations and rewards). Then, the goal of the agent is to find the actions that maximize the rewards. As an example the network administrator can set a target policy, for instance the delay of a set of flows, then the agent acts on the SDN controller by changing the configuration and for each action receives a reward, which increases as the in-place policy gets closer to the target policy. Ultimately, the agent will learn the set of configuration updates that result in such target policy. Recently, deep reinforcement learning techniques have provided important breakthroughs in the AI field, notable examples are \cite{nature1,nature2}.

Please note that learning can also happen offline and applied online. In this context knowledge can be learned offline training a neural network with datasets of the behavior of a large set of networks, then the resulting model can be applied online.  

\end{itemize}

\paragraph{Machine Learning $\rightarrow$ Intent Language}

The \ac{KP} eases the transition between telemetry data collected by the analytics platform and control specific actions. Traditionally, a network operator had to examine the metrics collected from network measurements and make a decision on how to act on the network. In \ac{KDN}, this process is partially offloaded to the \ac{KP}, which is able to make -or recommend- control decisions taking advantage of \ac{ML} techniques. The \ac{KP} expresses control decisions via an Intent-driven language (see Section~III-d), which eases the transition between the high-level decisions made by the \ac{KP} and the imperative language used by data plane elements.

Depending on whether the network operator is involved or not in the decision making process, there are two different sets of applications for the \ac{KP}. We next describe these potential applications and summarize them in Table~\ref{table:apps}.







\begin{table}[]
\centering
\caption{KDN applications}
\label{table:apps}
\begin{tabular}{r|l|l|l|}
\cline{2-4}
                                           & \textbf{Supervised}                                               & \textbf{Unsupervised} & \textbf{Reinforcement}                                            \\ \hline
\multicolumn{1}{|r|}{\textbf{Closed Loop}} & \begin{tabular}[c]{@{}l@{}}Automation\\ Optimization\end{tabular} & Improvement           & \begin{tabular}[c]{@{}l@{}}Automation\\ Optimization\end{tabular} \\ \hline
\multicolumn{1}{|r|}{\textbf{Open Loop}}   & \begin{tabular}[c]{@{}l@{}}Validation\\ Estimation\\What-if analysis\end{tabular}   & Recommendation        & N/A                                                               \\ \hline
\end{tabular}
\end{table}

\begin{itemize}
\item \textit{Closed loop:} When using supervised or reinforcement learning, the network model obtained can be used first for \textit{automation}, since the \ac{KP} can make decisions automatically on behalf of the network operator. Second, it can be used for \textit{optimization} of the existing network configuration, given that the learned network model can be explored through common optimization techniques to find (quasi)optimal configurations. Both applications can be also achieved using reinforcement learning, please note that with some techniques this can be achieved online and model-free. In the case of unsupervised learning, the knowledge discovered can be used to \emph{improve} automatically the network via the intent interface offered by the \ac{SDN} controller, although this requires additional research efforts. 

\item \textit{Open loop:} In this case the network operator is still in charge of making the decisions, however it can rely on the \ac{KP} to ease this task. When using supervised learning, the model learned by \ac{ML} can be used for \textit{validation}. In this case, the network administrator can query the model to validate the tentative changes to the configuration before applying them to the system. The model can also be used as a tool for performance \textit{estimation} and \textit{what-if analysis}, since the operator can tune the variables considered in the model and obtain an assessment of the network performance. 
When using unsupervised learning, the correlations found in the explored data may serve to provide \emph{recommendations} that the network operator can take into consideration when making decisions.

\end{itemize}

\paragraph{Intent Language $\rightarrow$ SDN controller}
\label{intent}

Both the network operators and the automatic systems that make decisions on their behalf express their intentions towards the network in the form of a declarative language. This serves to offer a common interface to both human and automatic decisions makers and to define precisely how the abstract intent should be translated into specific control directives. In contrast, the communication between the \ac{SDN} controller and data plane devices is done using imperative languages.

Using declarative languages for the \ac{SDN} northbound interface has been widely discussed in the \ac{SDN} literature \cite{frenetic,procera}. Among these declarative languages, the so-called Intent-driven ones (e.g. NEMO\footnote{http://www.nemo-project.net/}, GBP\footnote{https://wiki.openstack.org/wiki/GroupBasedPolicy}) are gaining traction in the industry at the time of this writing. Such languages allow abstract network directives to be rendered into specific control actions.

In that sense, the \ac{SDN} controller receives the declarative primitives through its northbound interface and then renders the Intent-driven language into specific imperative control actions. This is possible since it has a complete and global view of the network and can actuate on all the network devices from a centralized point. That way, it can find the appropriate control instructions to apply that reflect the expressed intent. At the time of this writing, there are already controllers (e.g., OpenDaylight\footnote{https://www.opendaylight.org/}) able to render some Intent-based languages such GBP. 

\paragraph{SDN controller $\rightarrow$ Forwarding Elements}

The parsed control actions are pushed to the forwarding devices via the controller southbound protocols in order to program the data plane according to the decisions made at the \ac{KP}. The controller may as well rely on management protocols (e.g. NETCONF) to reconfigure the data plane devices, if necessary. The forwarding elements at the data plane operate now based on the updated operational state and configuration pushed by the \ac{SDN} controller.

\section{Use-cases}
\label{sec:usecases}

This section presents a set of specific uses-cases that illustrate the potential applications of the \ac{KDN} paradigm and the benefits a \ac{KP} based on \ac{ML} may bring to common networking problems. For two representative use-cases, we also provide early experimental results that show the technical feasibility of the proposed paradigm. All the datasets used in this paper can be found at \cite{kdndataset}.


\subsection{Routing in an Overlay Network}
\label{sec:usecases:overlay}

The main objective of this use-case is to show that it is possible to model the behavior of a network with the use of \ac{ML} techniques. In particular, we present a simple proof-of-concept example in the context of overlay networks, where an \ac{ANN} is used to build a model of the delay of the (hidden) underlay network, which can later be used to improve routing in the overlay network.

Overlay networks have become a common solution for deployments where one network (overlay) has to be instantiated on top of another (underlay). This may be the case when a physically distributed system needs to behave as a whole while relying on a transit network, for instance a company with geo-distributed branches that connects them through the Internet. Another case is when a network has to send traffic through another for which it is not interoperable, for example when trying to send Ethernet frames over an IP-only network.

\begin{figure}
  \centering
  \includegraphics[width=0.99\columnwidth]{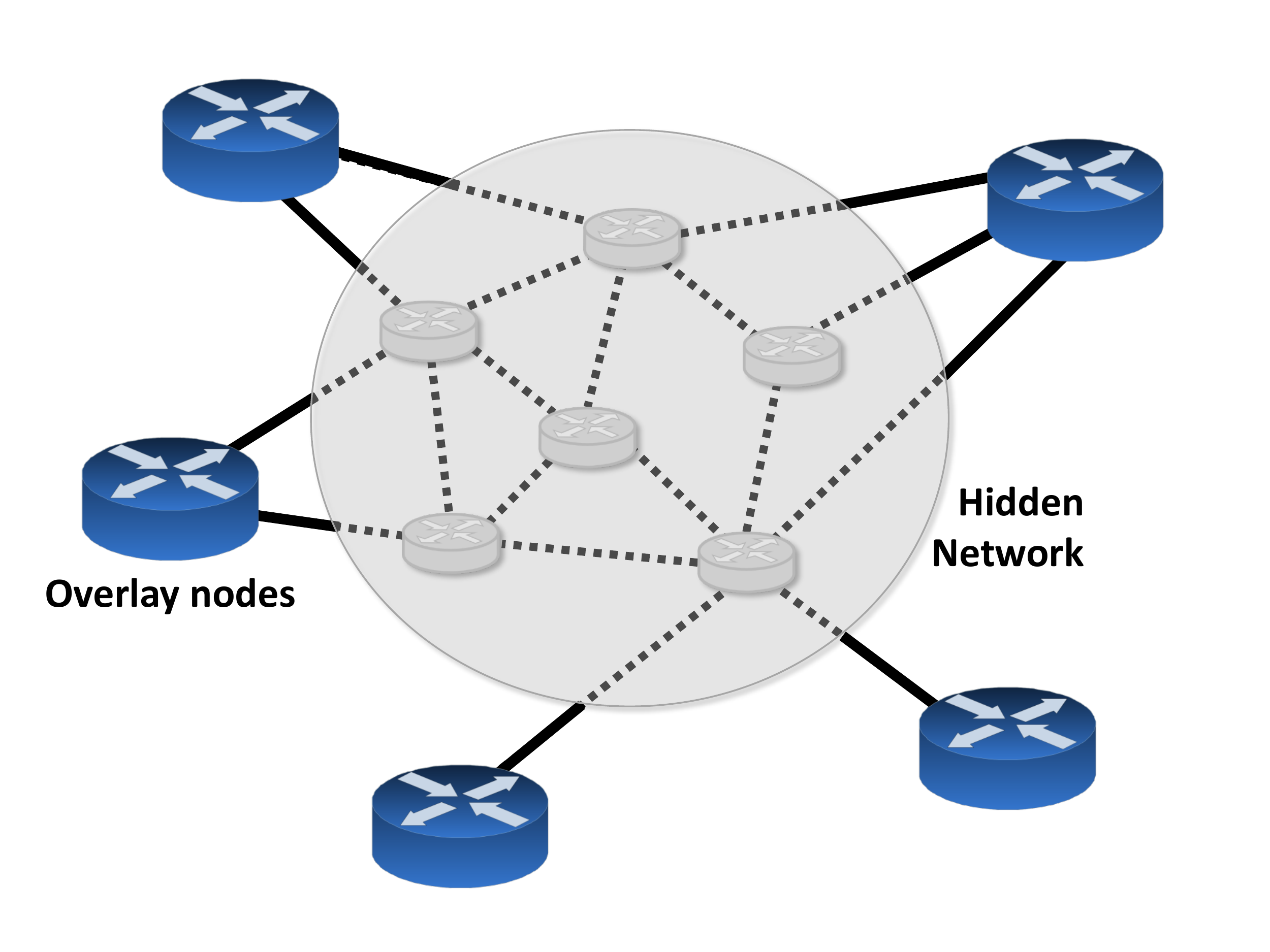}
  \caption{Overlay network with a hidden underlay}
  \label{fig:unetwork}
\end{figure}

In such cases, an overlay network can be instantiated by means of deploying overlay-enabler nodes at the edge of the transit network and then tunneling overlay traffic using an encapsulation protocol (e.g. LISP (RFC 6830), VXLAN (RFC 7348), etc). In many overlay deployments the underlay network belongs to a different administrative domain and thus its details (e.g. topology, configuration) are hidden to the overlay network administrator (see Fig.~\ref{fig:unetwork}).

Typically, overlay edge nodes are connected to the underlay network via several links. Even though edge nodes have no control over the underlay routing, they can distribute the traffic among the different links they use to connect to it. Edge nodes can use overlay control plane protocols (e.g. LISP)~\cite{arnatal} to coordinate traffic balancing policies across links. However, a common problem is how to find best/optimum per-link policies such that the global performance is optimized. An efficient use of edge nodes links is critical since it is the only way the overlay operator can control --to a certain extent-- the traffic path over the underlay network.

Overlay operators can rely on building a model of the underlay network to optimize the performance. However, building such a model poses two main challenges. First, neither the topology nor the configuration (e.g., routing policy) of the underlay network are known, and thus it is difficult to determine the path that each flow will follow. Second, mathematical or theoretical models may fall short to model such a complex scenario.

\ac{ML} techniques allow modeling hidden systems by analyzing the correlation of inputs and outputs in the system. In other words, \ac{ML} techniques can model the hidden underlay network by means of observing how the output traffic behaves for a given input traffic (i.e., \textit{f}(routing policy, traffic)~=~performance). For instance, if two edge node links share a transit node within the -hidden- underlay network, \ac{ML} techniques can learn that the performance decreases when both of those links are used at the same time and therefore recommend traffic balancing policies that avoid using both links simultaneously.

\subsubsection{Experimental Results}
To assess the validity of this approach we carried out the following simple experiment. We have simulated a network with 12 overlay nodes, 19 underlay elements and a total of 72 links. 
From the \ac{KP} perspective, only the overlay nodes that send and receive traffic are seen, while the underlay network is hidden. The network is simulated using Omnet++\footnote{https://omnetpp.org/} with the following characteristics: overlay nodes randomly split the traffic independently of the destination node, the underlay network uses shortest path routing with constant link capacity, constant propagation delay and Poisson traffic generation.

We train an \ac{ANN} using Pylearn2--Theano 0.7. The \ac{ANN} has one hidden layer, with a sigmoid activation function, and it is trained using the following input features: the amount of traffic among pairs of the overlay nodes and the ratio of traffic that is sent to each link. The average delays among paths obtained in the simulation are used as output features. Please note that the \ac{ANN} does not have access to any information regarding the underlay. We train the network with 9,600 training samples and we use 300 –-separate samples-- to validate the results. 

\begin{figure}
  \centering
  \includegraphics[width=0.90\columnwidth]{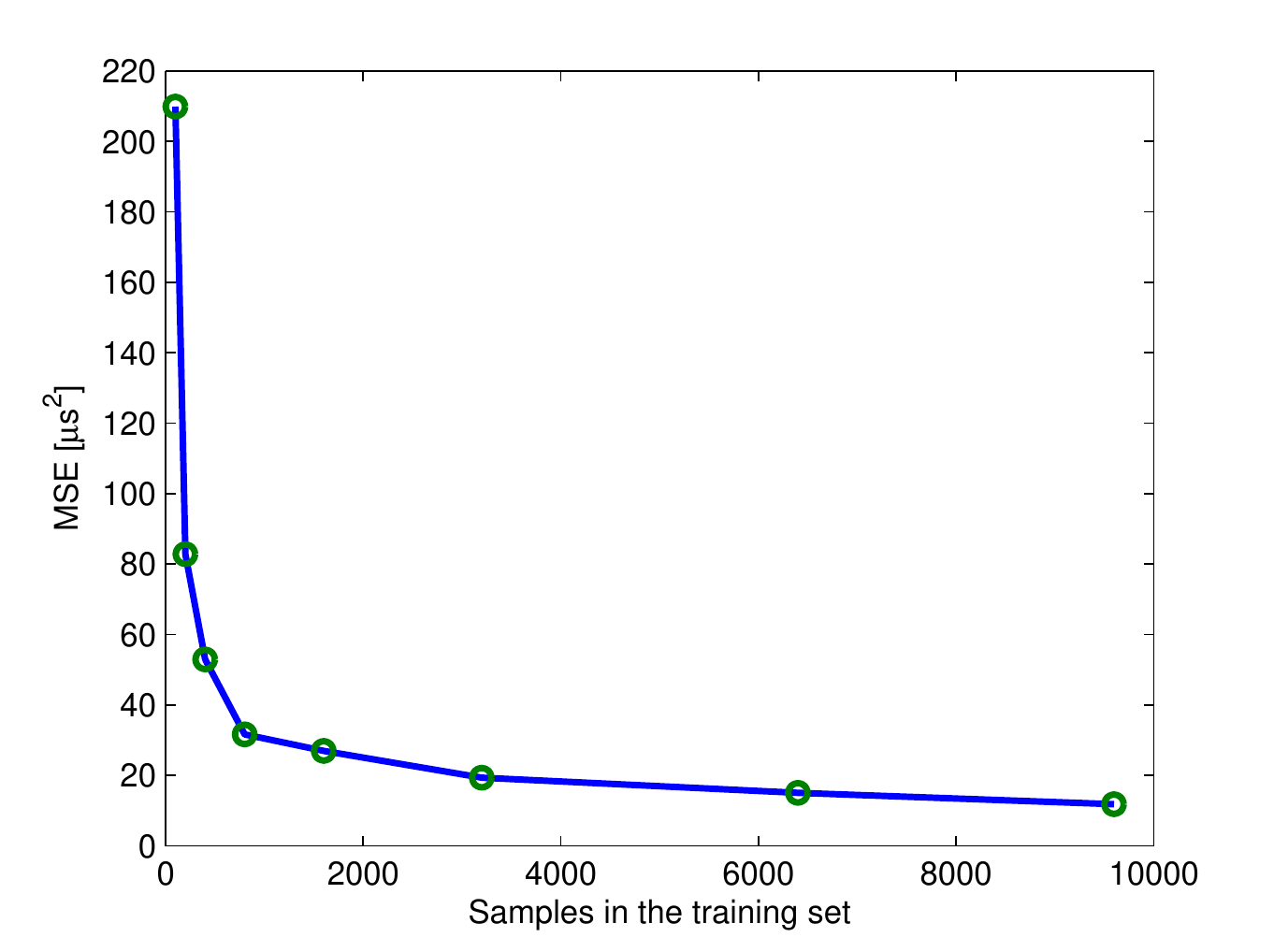}
  \caption{Evolution of the MSE (Mean Squared Error) as a function of the size of the training set.}
  \label{fig:TSsize}
\end{figure}

With this use-case we aim to learn the function that relates the traffic and the routing configuration of the overlay network with the resulting average delay of each path. The results show that the accuracy of the model is reasonably high, with a relative error of roughly 1\% when using 3,000 training samples. This error is computed as the average relative error of the delay of all paths in each of the 300 samples of the test set. Fig.~\ref{fig:TSsize} shows the accuracy (Mean Squared Error) as a function of the size of the training data set. The figure shows a typical exponential delay commonly found in \ac{ML}.    

\subsection{Resource Management in an NFV scenario}
\label{sec:usecases:nfv}

This use-case shows how the \ac{KDN} paradigm can also be useful in the context of \ac{NFV}.
\ac{NFV} \cite{nfv} is a networking paradigm where network functions (e.g., firewalls, load-balancers, etc.) no longer require specific hardware appliances but rather are implemented in the form of \acp{VNF} that run on top of general purpose hardware. 

The resource management in \ac{NFV} scenarios is a complex problem since \ac{VNF} placement may have an important impact on the overall system performance. The problem of optimal \ac{VM} placement has been widely studied for \ac{DC} scenarios (see~\cite{meng} and the references therein), where the network topology is mostly static. However, in \ac{NFV} scenarios the placement of a \ac{VNF} modifies the performance of the virtualized network. This increases the complexity of the optimal placement of VNFs in \ac{NFV} deployments.

Contrary to the overlay case, in the \ac{VNF} placement problem all the information is available, e.g. virtual network topology, CPU/memory usage, energy consumption, \ac{VNF} implementation, traffic characteristics, current configuration, etc. However, in this case the challenge is not the lack of information but rather its complexity. The behavior of VNFs depend on many different factors and thus developing accurate models is challenging.

The \ac{KDN} paradigm can address many of the challenges posed by the \ac{NFV} resource-allocation problem. For example, the \ac{KP} can characterize, via \ac{ML} techniques, the behavior of a \ac{VNF} as a function of the collected analytics, such as the traffic processed by the \ac{VNF} or the configuration pushed by the controller. With this model, the resource requirements of a \ac{VNF} can be modeled by the \ac{KP} without having to modify the network. This is helpful to optimize the placement of this \ac{VNF} and, therefore, to optimize the performance of the overall network. 

\subsubsection{Experimental results}
To validate this use-case we model the CPU consumption of real-world VNFs when operating under real traffic. We have chosen two different network pieces, an Open Virtual Switch (OVS v2.0.2\footnote{http://openvswitch.org/}) and Snort (v2.9.6.0\footnote{https://www.snort.org/}). We have tested OVS with two different set of rules and controller configurations: as a SDN-enabled firewall and as a SDN-enabled switch. In both cases, we have aimed to have a representative configuration of real-world deployments. 

To measure the CPU consumption of both VNFs we have deployed them in \acp{VM} (Ubuntu 14.04.1 running on top of VMware ESXi v5.5). The VNFs are virtually connected (using gigabit links) to two \acp{VM} that generate and receive traffic. The traffic used in this experiment was replayed using tcpreplay (version 3.4.4) from an on-campus DPI infrastructure. The campus network serves around 30k users. Details about the traffic traces can be found in~\cite{barlet}. To represent the traffic, we extract off-line a set of 86 traffic features in 20 second batches: number of packets, number of 5-tuple flows, average length, number of different IPs or ports, application layer information, among others. In the learning process, we use the Matlab \ac{ANN} toolbox with one hidden layer, where the input are all the traffic features, and the output is the measured CPU consumption. In this case we aim to learn the function that relates the traffic features with the CPU consumption.

\begin{figure}
  \centering
  \includegraphics[width=0.99\columnwidth]{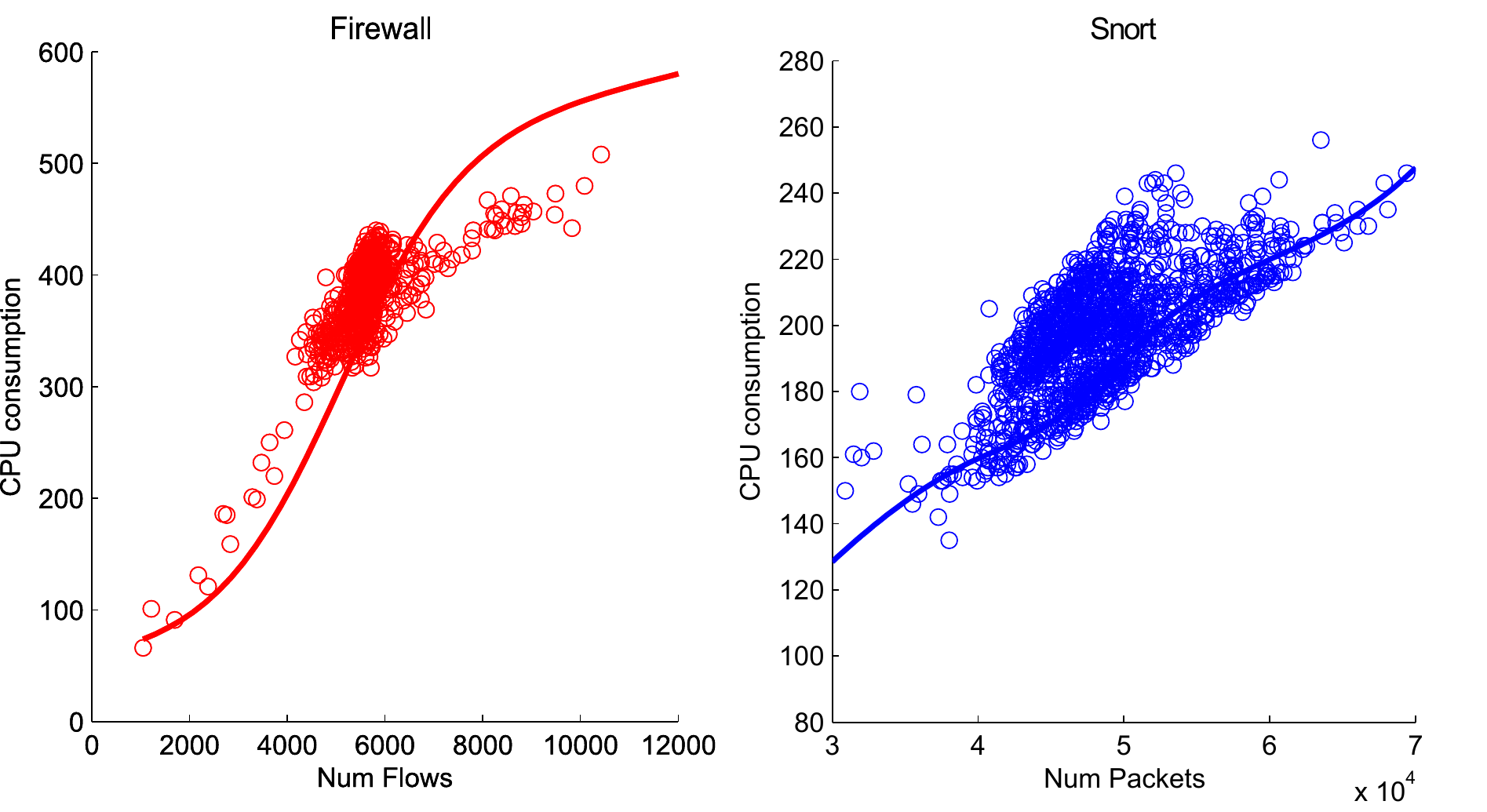}
  \caption{Measured points and the built model using two different features for two different VNF (only showing the most relevant feature)}
  \label{fig:vnfResults}
\end{figure}

\begin{figure}
  \centering
  \includegraphics[width=0.99\columnwidth]{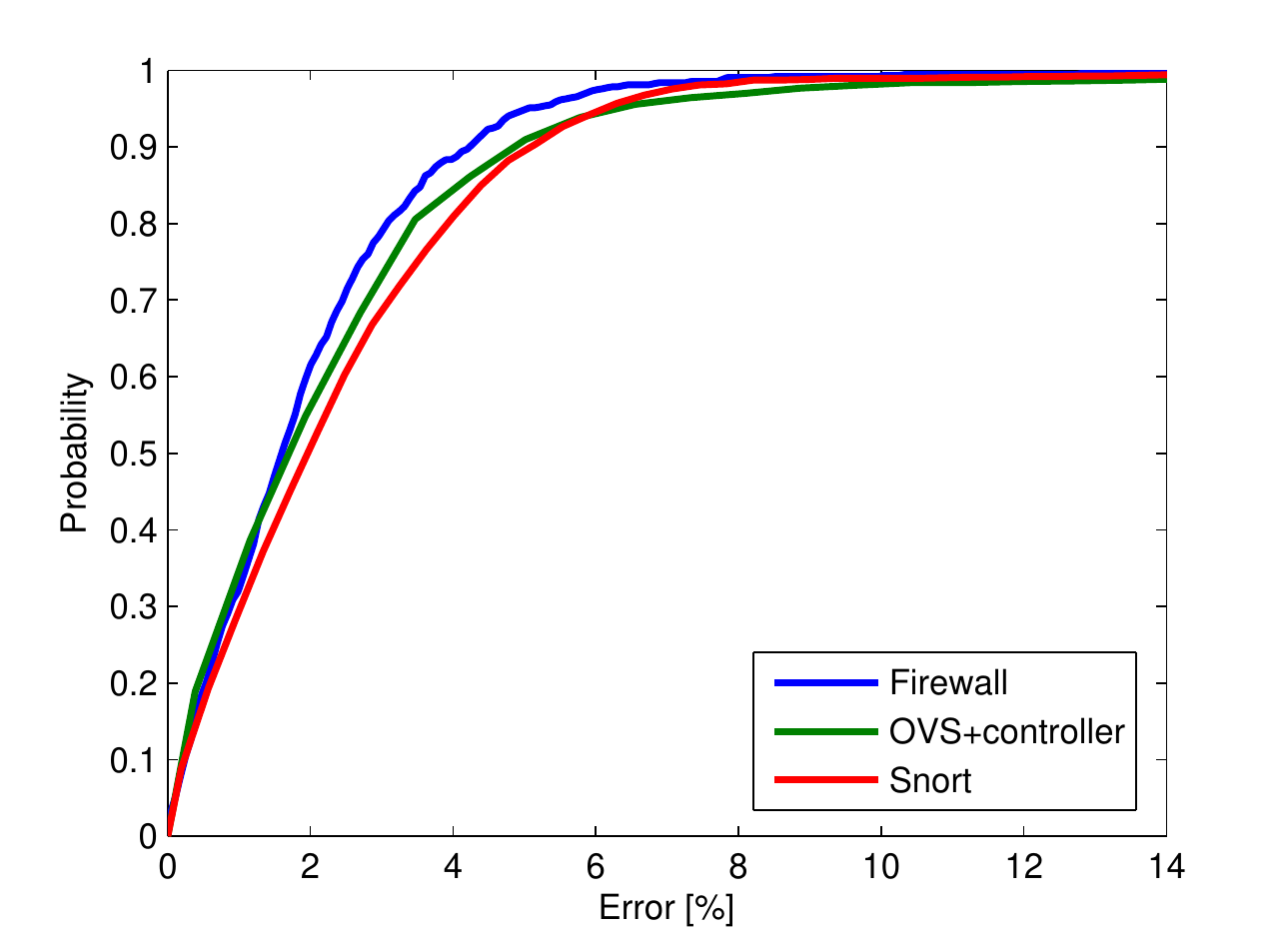}
  \caption{Cumulative Distribution Function of the relative error in the three VNFs}
  \label{fig:errorCDF}
\end{figure}

We train the \ac{ANN} with 600 samples for the OVS firewall model and 900 samples for Snort and the OVS switch. To assess the validity of the model we use different samples as a test set, 150 and 200 samples respectively. To show the complexity of the model and the need to use \ac{ML}, we first present the results when only one feature is used for prediction. Fig.~\ref{fig:vnfResults} shows that the model is non-linear and requires a multi-dimensional model. In particular, the figure plots the predicted CPU consumption (line) and the measured data (dots) as a function of the traffic feature used for prediction. The number of flows is used as predictor for the firewall, while the number of packets is used for Snort. Both figures show that, when only selecting one feature, the model is not accurate and non-linear, which motivates the use of \ac{ML}. Fig.~\ref{fig:errorCDF} shows a CDF of the relative error of the full model when trained with all the features, demonstrating that the model achieves reasonable accuracy.

\subsection{Knowledge extraction from network logs}
\label{sec:usecases:logs}

Operators typically equip their networks with a logging infrastructure where network devices report events (e.g., link going down, packet losses, etc.). Such logs are extensively used by operators to monitor the health of the network and to troubleshoot issues. Log analysis is a well-known research field (e.g., see \cite{logs} and the references therein) and, in the context of the KDN paradigm, it can also be used in networking. By means of unsupervised learning techniques, a KDN architecture can correlate log events and discover new knowledge, which can be used by the network administrators for network operation using the open-loop approach, or that can be handled automatically in a closed-loop solution via the Intent interface offered by the \ac{SDN} controller. The following table shows some specific examples:

\begin{table}[h]
\centering
\caption{Examples of Knowledge Discovery using Network Logging and unsupervised learning.}
\label{my-label}
\begin{tabular}{|l|}
\hline
\begin{tabular}[c]{@{}l@{}}Node N is always congested around 8pm and \\ Services X and Y have an above-average number of clients\end{tabular} \\ \hline
\begin{tabular}[c]{@{}l@{}}Abnormal number of BGP UPDATES messages sent\\ and Interface 3 is flapping\end{tabular}                            \\ \hline
\begin{tabular}[c]{@{}l@{}}Fan speeds increase in node N with frequency Y\\ Optics in interface Y fail\end{tabular}                            \\ \hline
\end{tabular}
\end{table}

\subsection{Short and long-term network planning}
\label{sec:usecases:planning}
Over time, network deployments typically have to face an increment in traffic load (e.g., higher throughput) and service requirements (e.g., less latency, less jitter, etc). Network operators have to deal with such increments and prepare the network in advance, in a process usually known as network planning. Network planning includes designing the network topology, selecting the specifications of the network hardware and deciding the traffic policies that distribute the traffic over the network. The objective of network planning is that in the long run the network meets the requirements of the network operator (and its subscribers, if any), that is to plan ahead to prevent potential bottlenecks, packet losses or performance drops \cite{qosAndPlanning}.

Network planning techniques commonly rely on computer models managed by experts that estimate the network capacity and forecast future requirements. Since this process is prone to errors, network planning typically results in over-provisioning. A KDN architecture can develop an accurate network model based on the historical data stored in the analytics platform. As a simple example, KDN can learn the relation between the number of clients (or the number of services) and the load and thus, accurately estimate when a network upgrade is required.

\section{Challenges and Discussion}
\label{sec:challenges}
The KDN paradigm brings significant advantages to networking, but at the same time it also introduces important challenges that need to be addressed. In what follows we discuss the most relevant ones.

\textit{New ML mechanisms:} Although \ac{ML} techniques provide flexible tools to computer learning, its evolution is partially driven by existing \ac{ML} applications (e.g., Computer Vision, recommendation systems, etc.). In this context the KDN paradigm represents a new application for \ac{ML} and as such, requires either adapting existing \ac{ML} mechanisms or developing new ones. A notable example are graphs, in networking, graphs are used to represent topologies, a fundamental part of the performance and features of a network. In this context, only preliminary attempts have been proposed in the literature to create sound ML algorithms able to model the topology of systems that can be represented through a graph \cite{graph1,graph2}. Although such proposals are not tailored to network topologies, their core ideas are encouraging for the computer networks research area. In this sense, the combination of ML techniques such as Q-learning techniques, convolutional neural networks and other deep learning techniques may be essential to make a step further in this area.

\textit{Non-deterministic networks:}
Typically networks operate with deterministic protocols. In addition, common analytical models used in networking have an estimation accuracy and are based on assumptions that are well understood. In contrast, models produced by techniques do not provide such guarantees and are difficult to understand by humans. This also means that manual verification is usually impractical when using models. Nevertheless, 
\ac{ML} models work well when the training set is \emph{representative} enough. Then, what is a \emph{representative} training set in networking? This is an important research question that needs to be addressed. Basically, we need a deep understanding of the relationship between the accuracy of the \ac{ML} models, the characteristics of the network, and the size of the training set. This might be challenging in this context as the \ac{KP} may not observe all possible network conditions and configurations during its normal operation. As a result, in some use-cases a training phase that tests the network under various representative configurations can be required. In this scenario, it is necessary to analyze the characteristics of such loads and configurations in order to address questions such as: does the normal traffic variability occurring in networks produce a representative training set? Does \ac{ML} require testing the network under a set of configurations that may render it unusable?

\textit{New skill set and mindset:} Networking started as a hardware-centric field of engineering, where pioneers designed and built hardware routers and switches. Since then, a new set of software engineering skills have become increasingly important. At the time of this writing, network devices already incorporate sophisticated pieces of software. With the rise of the \ac{SDN} paradigm, software development has become even more important in networking. This has created an important shift on the required expertise of networking engineers and researchers. The KDN paradigm further exacerbates this issue, as it requires a new set of skills: \ac{ML} techniques and, in general, knowledge of Artificial Intelligence tools. This
represents an important change in the mindset of the people working both in the industry and academia. 

\textit{Standardized Datasets:} In many cases, progress in ML techniques heavily depends on the availability of standardized datasets. Such datasets are used to research, develop and benchmark new AI algorithms. Some researchers argue that the cultivation of high-quality training datasets is even more important that new algorithms, since focusing on the dataset rather than on the algorithm may be a more straightforward approach. The publication of datasets is already a common practice in several popular ML application, such as image recognition \cite{imagenet}. In this paper we advocate that we need similar initiatives for the computer network AI field, were public training sets from experimental networks are published and used for research and development. All datasets used in this paper are public and can be found at \cite{kdndataset}.

\section{Conclusions}
\label{sec:conclusions}

In this paper, we introduced the concept of Knowledge-Defined Networking (KDN) a novel paradigm that combines Software-Defined Networking, Network Analytics and Machine Learning to ultimately provide automated network control. We also presented a set of use-cases and preliminary experimental evidence that illustrate the feasibility and advantages of the proposed paradigm. 
Finally, we discussed some important challenges that need to be addressed before completely achieving the vision shared in this paper. We advocate that addressing such challenges requires a truly inter-disciplinary effort between the research fields of Artificial Intelligence, Network Science and Computer Networks.

\section*{Acknowledgment}

This work has been partially supported by the Spanish Ministry of Education under grant FPU2012/01137, by the Spanish Ministry of Economy and Competitiveness and EU FEDER under grant TEC2014-59583-C2-2-R, and by the Catalan Government under grant 2014SGR-1427. 


\end{document}